**Biological activity of a $SiO_2$-CaO-$P_2O_5$ sol-gel glass highlighted by PIXE-RBS methods**

J. Lao[1], J.M. Nedelec[2], Ph. Moretto[3], E. Jallot[1*].

[1]*Laboratoire de Physique Corpusculaire de Clermont-Ferrand CNRS/IN2P3 UMR 6533 Université Blaise Pascal – 24 avenue des Landais, 63177 Aubière Cedex, France.*

[2]*TransChiMiC, Laboratoire des Matériaux Inorganiques CNRS UMR 6002 Université Blaise Pascal & ENSCCF – 24 avenue des Landais, 63177 Aubière Cedex, France.*

[3]*Centre d'Etudes Nucléaires de Bordeaux Gradignan CNRS/IN2P3 UMR 5797 Université Bordeaux 1 – Chemin du Solarium, Le Haut-Vigneau, BP 120, 33175 Gradignan Cedex, France.*

**\*Corresponding author:**

Edouard Jallot

Laboratoire de Physique Corpusculaire de Clermont-Ferrand CNRS/IN2P3 UMR 6533

Université Blaise Pascal – 24 avenue des Landais, 63177 Aubière Cedex, France.

**Tel** : 33 (0)4 73 40 72 65

**Fax** : 33 (0)4 73 26 45 98

**E-mail** : jallot@clermont.in2p3.fr

*Abstract*

It is proposed in this study to observe the influence of $P_2O_5$ on the formation of the apatite-like layer in a bioactive glass via a complete PIXE characterization. A glass in the $SiO_2$-CaO-$P_2O_5$ ternary system was elaborated by sol-gel processing. Glass samples were soaked in biological fluids for periods up to 10 days. The surface changes were characterized using Particle Induced X-ray Emission (PIXE) associated to Rutherford Backscattering Spectroscopy (RBS), which are efficient methods for multielemental analysis. Elemental maps of major and trace elements were obtained at a micrometer scale and revealed the bone bonding ability of the material. The formation of a calcium phosphate-rich layer containing magnesium occurs after a few days of interaction. We demonstrate that the presence of phosphorus in the material has an impact on the development and the formation rate of the bone-like apatite layer. Indeed, the Ca/P atomic ratio at the glass/biological fluids interface is closer to the nominal value of pure apatite compared to $P_2O_5$-free glasses. It would permit, *in vivo*, an improved chemical bond between the biomaterials and bone.

*Introduction*

With the development of biologically active materials, new field of applications arise in surgical therapeutics. Clinical operations on bone defects and fractures may call for a filling material that also presents the ability to contribute to the healing process. For this purpose, bioactive glasses are of huge interest. In contact with living tissues, bioactive glasses establish an enduring interface consisting of a calcium phosphate-rich layer that shows a bone-like apatite structure. The bioactivity mechanisms and growth of the layer at the interface deeply depend on the composition of the glass. The biological activity of bioactive glasses is linked to their capability, in aqueous solution, to leach ions from their surface; a porous silica-gel layer is then formed, which will play the part of support for bone-like apatite crystals growth [1]. The development mechanisms of that calcium phosphate- layer lie in the diffusion of calcium and phosphorus ions from the glass and from the aqueous medium to the material surface [2]. As a result, controlling the surface reactions rates and kinetics is of major importance. Optimizing determining parameters such as the material chemical composition and textural properties is in keeping with the concern that the material will be used as an efficient implant, capable of forming a strong interfacial bond with host tissues and stimulating bone-cell proliferation [3].

For this purpose, a $SiO_2$–$CaO$–$P_2O_5$ bioactive glass was elaborated using the sol-gel method, which permits the synthesis of materials with higher purity and homogeneity at low processing temperature [4]. Samples of gel-glass powder and glass compacted discs were immersed in biological fluids for varying periods. As a network former, phosphorus was expected to influence the in vitro bioactivity via a modification of the dissolution kinetics [5]. Analyses of major, minor and trace elements present at the biomaterial/biological fluids interface were performed by particle-induced X-ray emission (PIXE) associated to Rutherford backscattering spectroscopy (RBS). Obtaining PIXE elemental maps at a micrometer scale permits the complete follow-up of the bone-like layer formation along with major and trace element quantification. It allows important evaluation for the in vivo bioactivity of such a bone-forming material.

*Materials and methods*

**Preparation of the bioactive glass samples**

Gel-glass powders containing 67.5wt% $SiO_2$-25wt% $CaO$-7.5 wt% $P_2O_5$ were prepared using the sol-gel process. Tetraethylorthosilicate (TEOS; $Si(OC_2H_5)_4$), triethylphosphate ($PO(OC_2H_5)_3$) and calcium nitrate $Ca(NO_3)_2 \cdot 4H_2O$ were mixed in a solution of ethanol in presence of water. The prepared sol was then transferred to an oven at 60°C for gelification and aging. Four hours later, the obtained gel was dried at 125°C for 24 hours, then finally reduced to powder and heated at 700°C for 24 hours. The final surface area of the glass was found to be 111 $m^2/g$ by nitrogen sorption analysis. Part of the dry gel powder was then compacted into discs of 13 mm diameter and 2 mm height.

**In vitro assays**

The glass discs were immersed at 37°C for 1, 6 h and 1, 2, 5, 10 days in 45 mL of a standard Dulbecco's Modified Eagle Medium (DMEM, Biochrom AG, Germany), which composition is almost equal to human plasma. 10 mg of gel-glass powder samples were soaked at 37°C for 1, 6 h and 1, 2, 3, 4 days in DMEM, with a surface area to DMEM volume ratio fixed at 500 $cm^{-1}$. After interaction, the samples were removed from the fluid, air dried and embedded in resin (AGAR, Essex, England). Before characterization, the glass discs were cut into thin sections of 30 micrometers nominal thickness using a Leica RM 2145 microtome. 1000 nm thin sections of the glass powder samples were prepared by mean of a Leica EM UC6 Ultramicrotome, and laid out on 50 mesh copper grids. Then, the sections and grids are placed on a mylar film with a hole of 3 mm in the centre.

**PIXE-RBS analysis**

Analyses of the biomaterial/biological fluids interface were carried out using nuclear microprobes at CENBG (Centre d'Études Nucléaires de Bordeaux-Gradignan, France). For PIXE analyses, we chose proton scanning micro-beam of 1.5 MeV energy and 100 pA in intensity. The beam diameter was nearly 2 μm. An 80 $mm^2$ Si(Li) detector was used for X-ray detection, orientated at 135° with respect to the incident beam axis and equipped with a beryllium window 12 μm thick. PIXE spectra are treated with the software package GUPIX [6]. Relating to RBS, a silicon particle detector placed 135° from the incident beam axis provided us with the number of protons that interacted with the sample. Data were treated with the SIMNRA code [7].

*Results and discussion*

**Glass powder samples**

Elemental maps for each immersion time in DMEM were recorded. Figure 1 represents the elemental distribution of a powder grain after 1 h of interaction with biological fluids. The grain is still homogeneous and dissolution has not begun. Its composition is in the order of the primary $SiO_2$-$CaO$-$P_2O_5$ synthesized glass. Nevertheless, some grains (not shown) present a gradient of Ca and P concentration from the centre to the periphery of the material, indicating that ionic exchanges are imminent. After 6 h soaking, we note that calcium and phosphorus started to diffuse from the glass (data not shown). Ion exchange between the grains and the solution has occurred and traces of magnesium are detected at the periphery of the material. However silicon is still uniformly distributed through the grains. The breakdown of the silicate network occurs within 24 h of interaction, and a calcium phosphate-rich layer is formed on particular nucleation sites, located at the periphery of the grains. Calcium and phosphorus ions continue to diffuse from the glass; those are added to the calcium ions and phosphates coming from biological fluids, forming an amorphous calcium phosphate layer on the glass surface. That is illustrated in Figure 2, which shows the multi-elemental maps of powder grains after 2 days soaking. As visible on the biggest grain, a homogeneous calcium phosphate-rich layer containing magnesium surrounds the material. The core of the grain is composed of the silicate network enduring dissolution. The smallest grains (in the picture corners) already changed into calcium phosphates.

**Glass compacted discs**

Glass pastilles react more slowly than powder samples, since their massive powder compacted shape does not grant the same porous-gel open structure as single grains. However, their retarded behavior is similar to that of powder grain samples. Figure 3 shows the multi-elemental maps across the periphery of a glass pastille after 1 h soaking. Measuring the elemental concentrations in the material reveals no changes in the material composition. On 6 h immersed samples, we observe the presence of thin Ca-P enriched areas disseminated on the surface of the discs. Growth of those areas is supplied by constant ionic exchanges between the material and biological fluids. It results in the formation of a large calcium phosphate layer after a few days of interaction (Figure 4). We have measured the Ca/P atomic ratio at the periphery of the glass: it is equal to 1.89 after 10 d soaking. That is an essential indication for the formation of bone-like apatite, which Ca/P nominal value is equal to 1.67.

*Conclusion*

Thanks to micro-PIXE associated to RBS, we are able to specify the role of major and trace elements in physico-chemical reactions occurring at the periphery of the glass. In contact with body fluids, bioactive glasses induce a specific biological response at their surface. The initial $SiO_2$–$CaO$–$P_2O_5$ glass network is quickly enduring dissolution. Then, following the different stages of the bioactivity process, a bone-like layer is quickly formed at the material periphery. The calcium phosphate-rich layer formation and evolution of the glass network are highlighted. Magnesium is proved to be blended into the material: that is new information of capital importance since magnesium can play an important role during spontaneous formation of in vivo calcium phosphates and bone bonding [8, 9]. The specific preparation protocol developed permits the characterization of highly porous powders with grains of a few micrometers.

We demonstrate that the presence of phosphorus in the material composition has an impact on the development and the formation rate of the bone-like apatite layer. In a previous work on $P_2O_5$-free glasses, we found that the Ca/P atomic ratio at the material periphery was equal to 2.05 after 10 d of interaction [10]. The Ca/P atomic ratio at the $SiO_2$–$CaO$–$P_2O_5$ glass/biological fluids interface is equal to 1.89 after 10 d soaking, which is closer to the 1.67 nominal value of pure apatite. Furthermore, phosphorus-based glass compacted discs present slower dissolution kinetics compared to $P_2O_5$-free glasses. It might permit, in vivo, an improved bonding ability with host tissues. Biological studies are now being performed to confirm this point.

*Figure captions*

Figure 1: Elemental maps of a $SiO_2$–$CaO$–$P_2O_5$ powder grain after 1 h of interaction with biological fluids (53 × 53 $\mu m^2$).

Figure 2: Elemental maps of $SiO_2$–$CaO$–$P_2O_5$ powder grains after 2 days of interaction with biological fluids (101 × 101 $\mu m^2$).

Figure 3: Elemental maps at the periphery of a $SiO_2$–$CaO$–$P_2O_5$ glass disc after 1 h of interaction with biological fluids (53 × 53 $\mu m^2$).

Figure 4: Elemental maps at the periphery of a $SiO_2$–$CaO$–$P_2O_5$ glass disc after 10 d of interaction with biological fluids (179 × 179 $\mu m^2$).

*Black-and-white figures (to be published)*

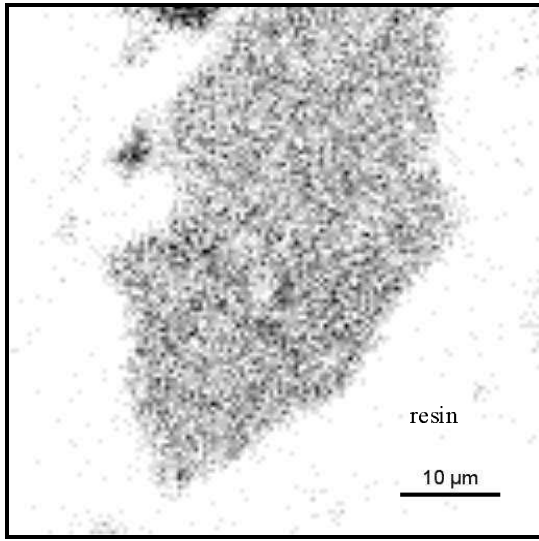

**Ca**

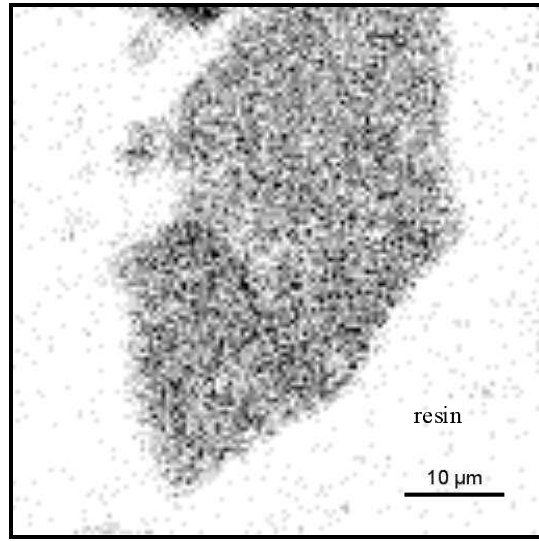

**P**

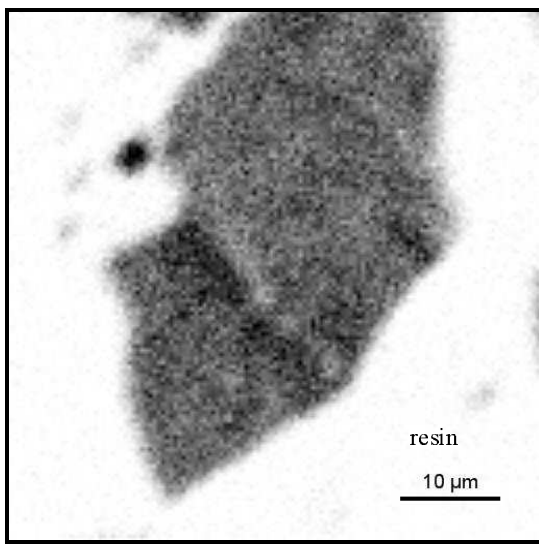

**Si**

Figure 1

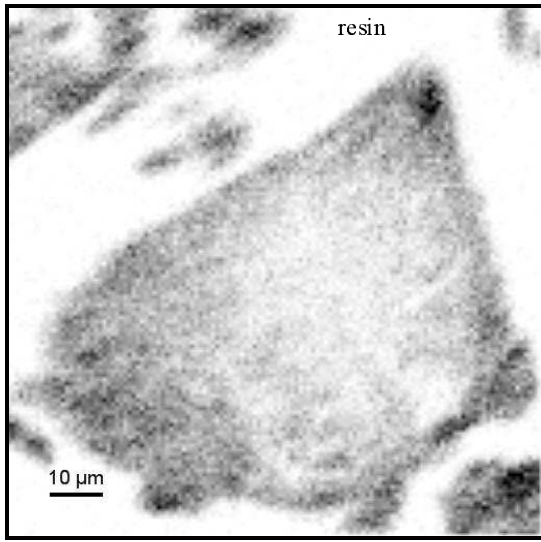 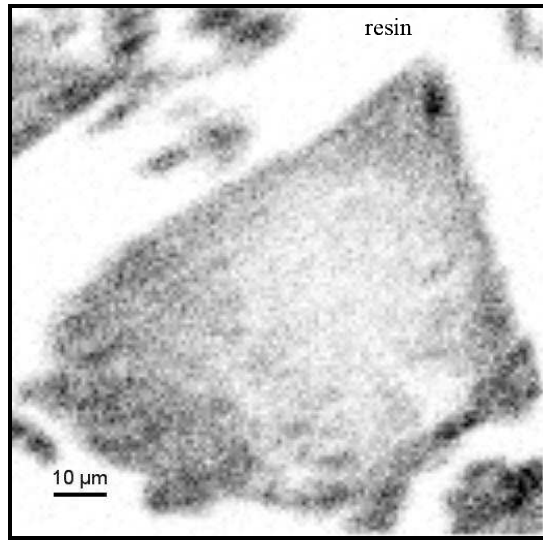

**Ca**        **P**

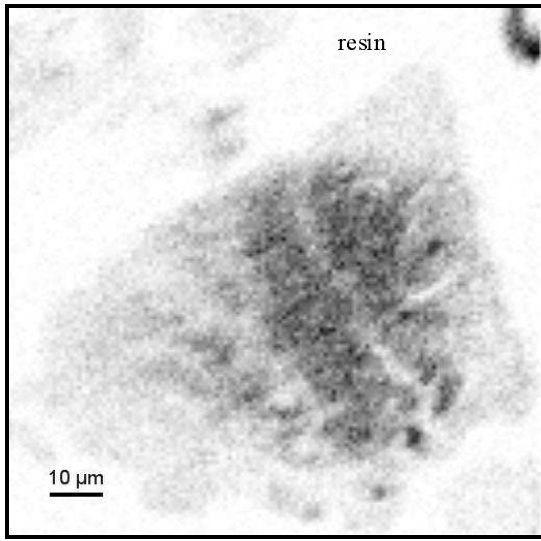 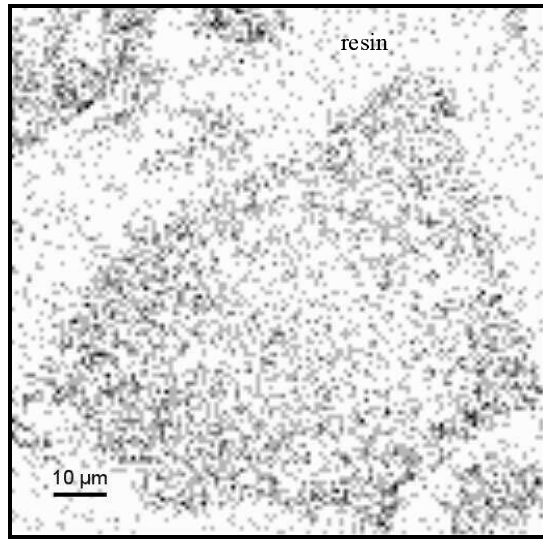

**Si**        **Mg**

Figure 2

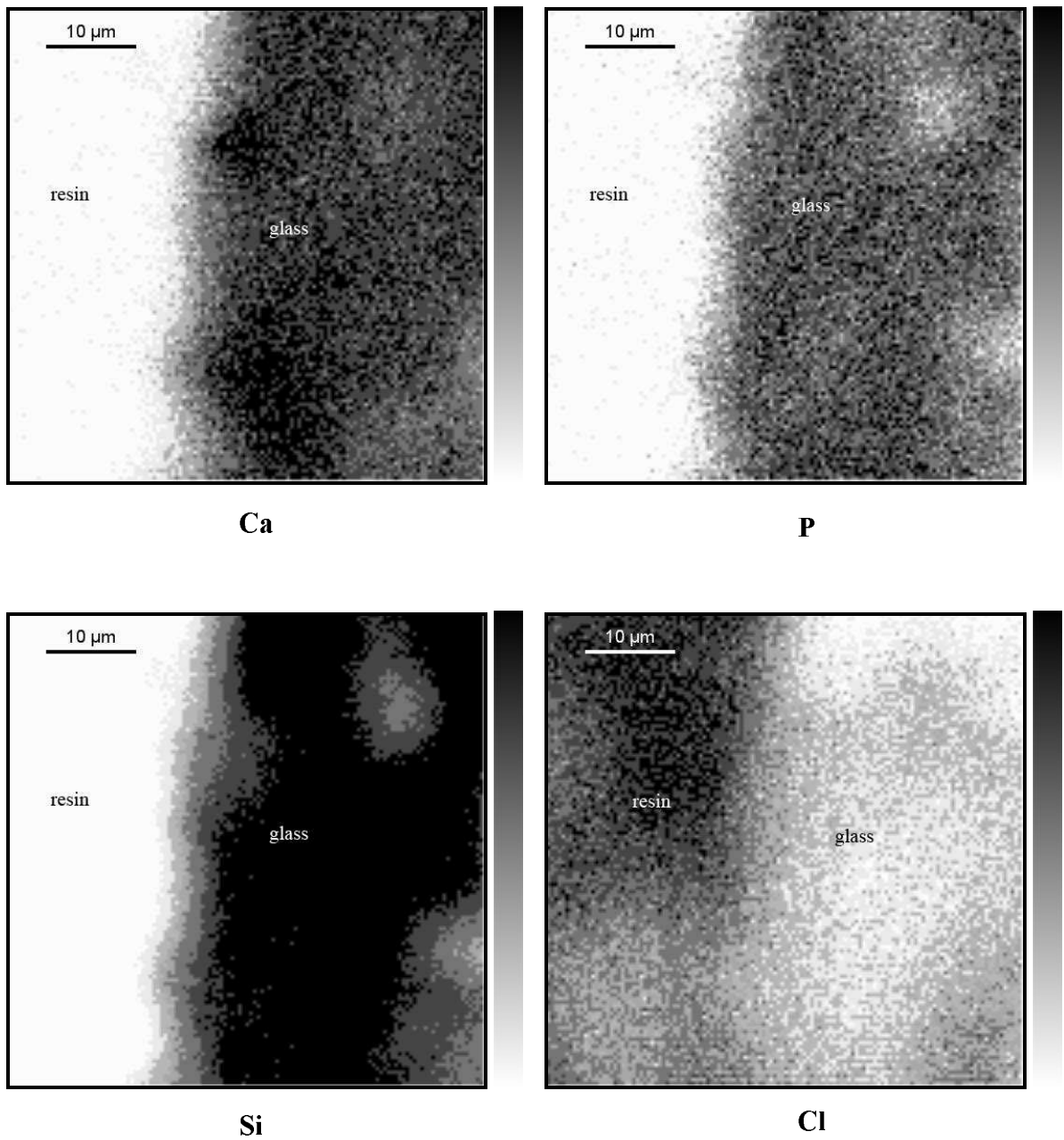

Figure 3

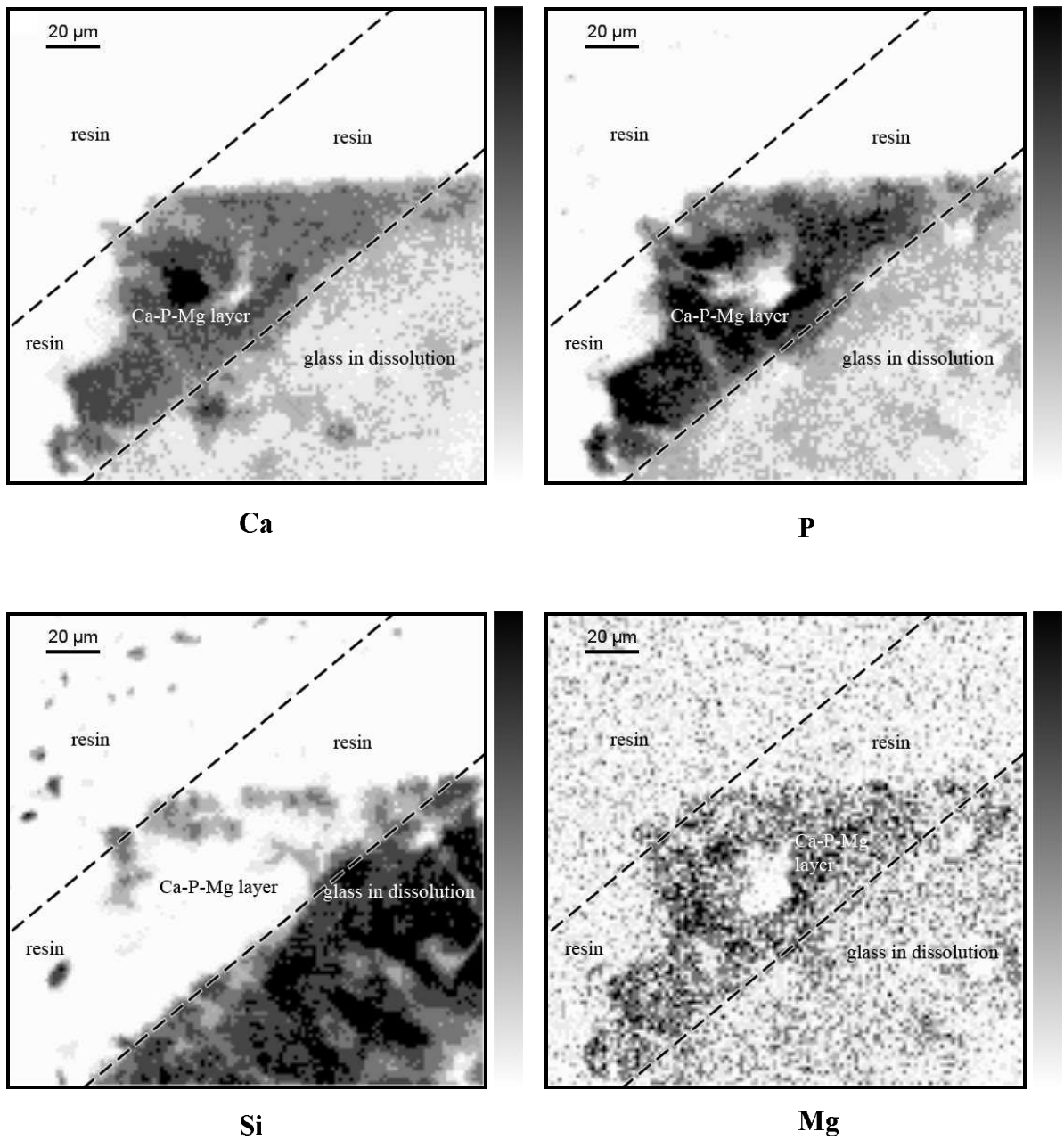

Figure 4